\documentclass[10pt,conference]{IEEEtran}
\RequirePackage{fix-cm}
\UseRawInputEncoding
\usepackage{inputenc}
\usepackage{graphicx}
\usepackage{tikz}
\usetikzlibrary{positioning}
\usetikzlibrary{trees,positioning,shapes,shadows}
\usepackage[edges]{forest}
\usepackage{tabularray}
\usepackage{forest}
\usepackage[normalem]{ulem}
\useunder{\uline}{\ul}{}
\usepackage{tabularray}
\usepackage{amsmath}
\usepackage{commath}
\usepackage{multirow}
\usepackage{graphicx}  
\usepackage{microtype}
\usepackage{tabularx}
\usepackage{makecell}
\usepackage{tabularray}
\usepackage{listings}
\usepackage{xcolor}
\usepackage{url}
\usepackage{graphicx} 
\usepackage{float}
\usepackage{caption}
\usepackage{listings}
\AtBeginDocument{%
  \providecommand\BibTeX{{%
    Bib\TeX}}}

\usepackage{hyperref}
\usepackage{cite}
\usepackage{amsmath,amssymb,amsfonts}
\usepackage{algorithmic}
\usepackage{graphicx}
\usepackage{textcomp}
\usepackage{xcolor}
\def\BibTeX{{\rm B\kern-.05em{\sc i\kern-.025em b}\kern-.08em
    T\kern-.1667em\lower.7ex\hbox{E}\kern-.125emX}}

\begin{document}

\title{CFCEval: Evaluating Security Aspects in Code Generated by Large Language Models}
\author{
\IEEEauthorblockN{Cheng Cheng}
\IEEEauthorblockA{Department of Computer Science and \\ Software Engineering, Concordia University,\\
Montreal, Canada\\
cheng.cheng.20171@mail.concordia.ca}
\and
\IEEEauthorblockN{Jinqiu Yang}
\IEEEauthorblockA{Department of Computer Science and \\ Software Engineering, Concordia University,\\
Montreal, Canada\\
jinqiu.yang@concordia.ca}
}
% \author{\IEEEauthorblockN{Cheng Cheng}
% \IEEEauthorblockA{\textit{Department of Computer Science and Software Engineering} \\
% \textit{Concordia University}\\
% Montreal, Canada \\
% cheng.cheng.20171@mail.concordia.ca}
% \and
% \IEEEauthorblockN{Jinqiu Yang}
% \IEEEauthorblockA{\textit{Department of Computer Science and Software Engineering} \\
% \textit{Concordia University}\\
% Montreal, Canada \\
% jinqiu.yang@concordia.ca}
% \and

\maketitle

\begin{abstract}
Code-focused Large Language Models (LLMs), such as CodeX and Star-Coder, have demonstrated remarkable capabilities in enhancing developer productivity through context-aware code generation. However, evaluating the quality and security of LLM-generated code remains a significant challenge. Existing evaluation protocols for Code LLMs lack both methodological rigor and comprehensive scope. A key limitation is dataset bias, which arises from unintentional overlap between training and testing data. Furthermore, while CodeBLEU, a BLEU-based metric, is widely used to assess code similarity, it suffers from critical shortcomings, including imprecise tokenization, structural limitations, and low reference diversity.
To address these challenges, we introduce CFCEval, a novel framework for evaluating the quality and security of code generated by LLMs. CFCEval mitigates dataset bias by creating a new benchmark, MLVBench, and incorporates ELRM, a new metric designed to assess the relevance between reference code and generated code. CFCEval evaluates generated code across four dimensions: programming quality, vulnerability-fixing capability, post-transformation fixing capability, and relevance. Our experiments show that CFCEval not only captures both quality and security aspects of generated code more effectively but also that its ELRM aligns more closely with human judgments than CodeBLEU, thus paving the way for future advancements in Code LLMs evaluation.
\end{abstract}


\section{Introduction}
\label{sec:intro}
Recent advancements in large language models (LLMs) ~\cite{deepseekai2025deepseekr1,schoenegger2025largelanguagemodel,openai2023gpt4} have significantly improved performance across various natural language processing tasks, including text generation, machine translation, and question answering. Building on these successes, domain-specific LLMs, or code-focused LLMs, have emerged to address programming-related tasks. Models such as DeepSeek~\cite{deepseekai2025deepseekr1}, CodeX~\cite{chen2021evaluating}, Code Llama ~\cite{roziere2024codeLlama}, and StarCoder2~\cite{lozhkov2024starcoder} excel at generating syntactically correct and semantically meaningful code, completing functions, and synthesizing full code blocks. Integrated into tools like GitHub Copilot and CodeGeeX, these models serve as intelligent coding assistants, enhancing developer efficiency, reducing repetitive tasks, and supporting rapid prototyping. Their widespread use marks a shift towards AI-augmented software engineering workflows.

Code-focused large language models (LLMs) are typically evaluated using benchmark datasets and metrics tailored to specific code-related tasks.
To assess the functional correctness of generated code, datasets such as HumanEval ~\cite{chen2021evaluating} and DS-1000~\cite{szalontai2024largelanguagemodelscode} are widely adopted, where models are evaluated based on whether their outputs pass a predefined set of test cases.
In addition, evaluation metrics such as Exact Match (EM)~\cite{ding2023crosscodeeval}, perfect accuracy, BLEU~\cite{bleu2022Papi}, and CodeBLEU~\cite{ren2020codebleu} are commonly used to quantify performance in tasks including text-to-code synthesis, code translation, and code change prediction~\cite{chen2021evaluating}.
These metrics primarily measure the degree of similarity between the generated code and the reference solution, typically at the token or n-gram level.
Among them, CodeBLEU extends BLEU by incorporating additional structural features, such as abstract syntax tree (AST) match and data flow consistency, and has demonstrated improved alignment with semantic correctness in code-related evaluations.

Despite the widespread adoption of datasets and evaluation metrics, current assessment methodologies for code-focused large language models (LLMs) suffer from critical limitations in both methodological rigor and scope. Specifically, in code completion tasks, especially from a security perspective, two major challenges persist: dataset bias and deficiencies in metrics related to code structure, semantic correctness, and prediction diversity.

Training-testing dataset overlap, particularly in benchmarks derived from public repositories like GitHub, inflates performance estimates due to memorization rather than true generalization. Existing metrics, such as CodeBLEU, fail to capture crucial aspects of code quality and security. These limitations include imprecise tokenization (e.g., treating code without whitespace as a single token), structural issues (e.g., short code fragments not forming valid Abstract Syntax Trees or data-flow graphs), and low reference diversity (e.g., relying on a single ground-truth reference despite multiple valid repair predictions). Consequently, these shortcomings hinder the ability to fully assess the quality and security of code generated by LLMs.
These limitations obscure the true capabilities of code LLMs and call for more principled, generalization-sensitive, and semantically aware evaluation frameworks.

To address these limitations, we introduce the Code Fix Capability Evaluation (CFCEval) framework, a novel methodology designed to assess both the quality and security of code LLMs. CFCEval mitigates training-testing data overlap bias by leveraging a new evaluation dataset, MLVBench, which introduces input perturbations through techniques such as variable renaming and structural refactoring. This enables the evaluation of model generalization on transformed inputs that remain semantically equivalent to the original ones.

The CFCEval framework evaluates code LLMs across four key dimensions: Programming Language Quality, Fixing Capability, Post-Transformation Fixing Capability, and Element-Level Relevance. These dimensions are designed to provide a comprehensive assessment of code generation models, considering both correctness and security aspects. To assess Element-Level Relevance, we introduce the Element-Level Relevance Metric (ELRM), which quantifies the degree of similarity between the generated code and the secure reference code. ELRM evaluates the relevance by analyzing keyword and operator matches, as well as string literal similarity, allowing for a fine-grained comparison of how well the generated code aligns with the intended secure solution. This metric enables a deeper understanding of the model’s performance, going beyond surface-level syntactic comparisons to consider the semantic accuracy of the generated fixes.

We perform extensive experiments to assess the effectiveness of the CFCEval framework and the Element-Level Relevance Metric (ELRM), and their correlation with evaluation scores from intelligent code LLMs on generated code. These experiments are designed to evaluate both the quality and security aspects of the code produced by the models, as well as the overall performance of the CFCEval framework in assessing these dimensions, addressing the research questions outlined in Section~\ref{experiment}. The results demonstrate that CFCEval, through the use of ELRM, effectively differentiates the performance of code LLMs, capturing both fine-grained code quality and the security aspects of the generated fixes. Furthermore, ELRM shows a stronger alignment with human-assigned LLM quality scores, outperforming commonly used evaluation metrics such as CodeBLEU and BLEU.
To further validate the robustness of CFCEval as an automatic evaluation benchmark, we also incorporate GPT-based metrics, providing additional support for the framework's ability to evaluate code generation in a comprehensive and reliable manner. In addition to these experiments, we present a pilot study in Section~\ref{study} that further validates the CFCEval framework's effectiveness in real-world vulnerability repair tasks, demonstrating its ability to assess code generation models in a practical context.

The paper is organized as follows. Section~\ref{framework} introduces the CFCEval framework, outlining its benchmark, evaluation dimensions, and the ELRM metric. Section~\ref{experiment} presents the experimental results, addressing the research questions and key findings. Section~\ref{limit} discusses the threats to validity, including limitations in the experimental setup and potential biases. Section~\ref{study} provides a pilot case study demonstrating the framework's application in a real-world context. Section~\ref{related_work} reviews related work, highlighting differences between existing methods and the proposed approach. Finally, Section~\ref{conclusion} summarizes the main findings and suggests directions for future research.

\begin{figure}[t]
\centering
\includegraphics[width=1\linewidth]{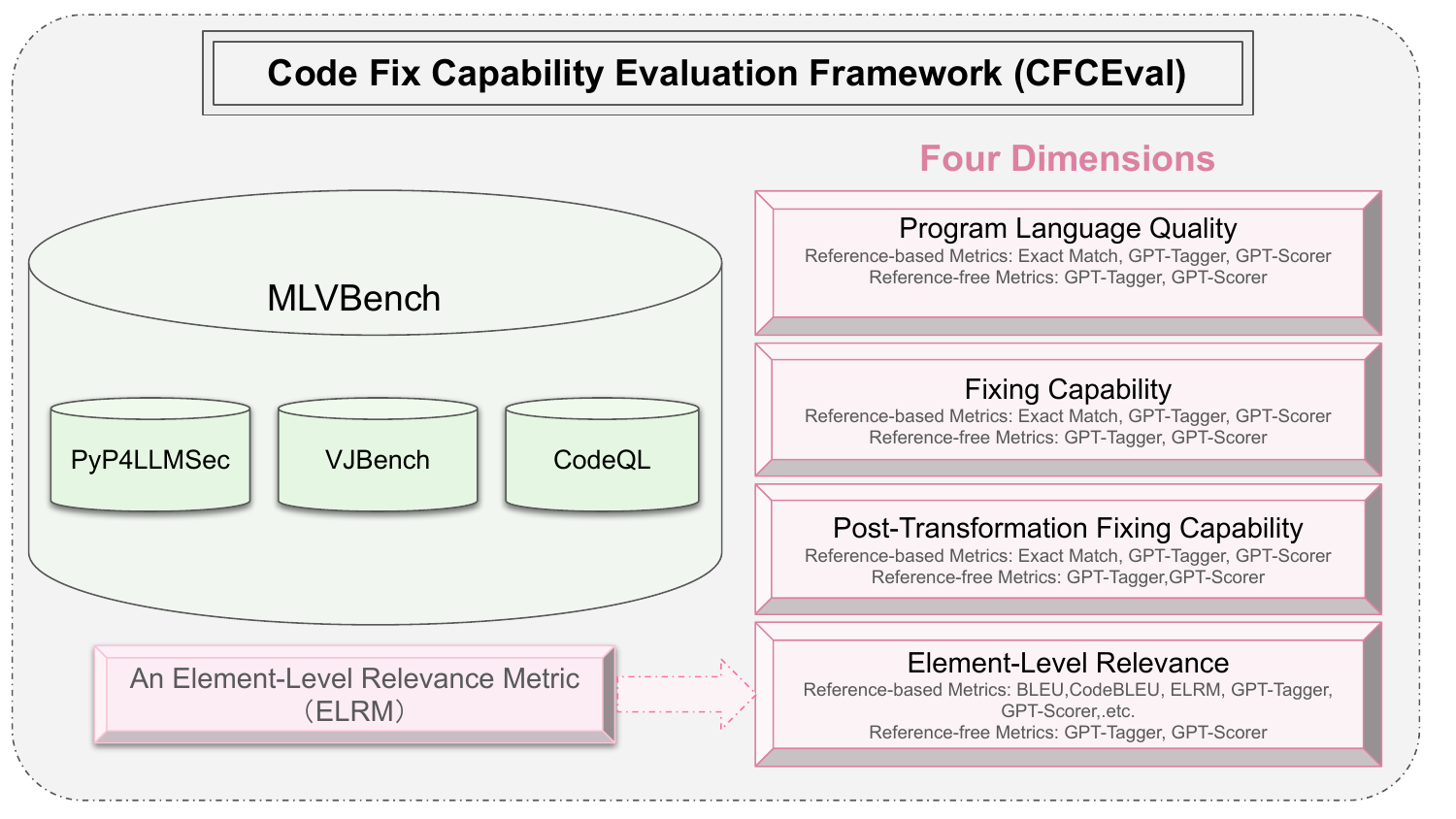}
\caption{Overview of the CFCEval framework, highlighting the dataset, the four evaluation dimensions, and the corresponding metrics used to assess the code quality and security of Code LLMs.}
\label{fig:framework}
\vspace{-1.5em}
\end{figure} 

\section{Code Fix Capability Evaluation Framework}
\label{framework}
To address the limitations discussed in Section~\ref{sec:intro}, namely, training-testing data overlap and the lack of meaningful evaluation metrics, we propose the Code Fix Capability Evaluation  (\textbf{CFCEval}) framework, a structured approach for evaluating the quality and security of Code LLMs.

CFCEval consists of three integral components:
(1) A bias-mitig-ating \textbf{dataset} created through semantic-preserving code transformations;
(2) A set of evaluation \textbf{dimensions} that collectively assess programming quality, vulnerability repairability, transformation robustness, and relevance;
(3) An evaluation metric, \textbf{ELRM}, built upon CodeBLEU and BLEU, designed to evaluate the relevance dimension.
Each component is introduced in turn in the following subsections.

\subsection{Dataset Construction for Bias Mitigation}
As discussed in the introduction, one critical limitation in existing evaluation protocols is the presence of dataset bias, which stems from unintentional overlap between training and testing data.
This issue is particularly pronounced in code LLMs, which are typically pre-trained on large-scale open-source repositories such as GitHub.
Many benchmark datasets are constructed from the same or similar codebases, often containing duplicated snippets, common templates, or boilerplate code structures.
Consequently, models may achieve artificially high scores by memorizing seen patterns, rather than demonstrating true generalization.
Such bias not only undermines the credibility of evaluation results but also obscures the model's ability to handle novel or obfuscated inputs-especially in security-critical settings.
To address this issue, our evaluation framework introduces a dedicated dataset, MLVBench, designed with controlled code perturbations (e.g., variable renaming and structural refactoring) to produce vulnerability instances that are semantically equivalent but distributionally shifted from typical training data.

\textbf{Dataset Construction.}
Several benchmarks have been developed for automated vulnerability repair, including Vul4J \cite{bui2022vul4j} for Java, Vul4C \cite{liu2025sok} for C/C++, and CVE-Bench \cite{wang-etal-2025-cve}, which covers multiple CVE-based samples. However, most of these datasets are language-specific, limiting their ability to evaluate the cross-language generalization of repair models.
To address this limitation, we construct a multi-language dataset, \textbf{MLVBench}, by unifying and adapting three public resources: PyP4LLMSec \cite{cheng2024pyp4llmsec}, VJBench \cite{Effective2023CodeTransform}, and CodeQL\footnote{\url{https://github.com/github/codeql}} security examples.
Specifically, VJBench includes 35 transformed Java vulnerabilities derived from Vul4J by applying systematic transformation strategies, such as identifier renaming and structural modifications. The same transformation methodology is further applied to selected samples from PyP4LLMSec and CodeQL security examples, ensuring consistency in data representation across programming languages.
PyP4LLMSec provides 156 vulnerabilities and 295 Python patches, covering 15 distinct CWE types \cite{cheng2024pyp4llmsec}.
CodeQL security examples offer a broad spectrum of vulnerabilities designed to demonstrate CodeQL query usage, spanning multiple CWE types across various programming languages. 
Inspired by the approach in prior work~\cite{pearce2021asleep}, we incorporate CodeQL into our benchmark to extract vulnerability patterns and facilitate dataset construction.
The distribution details of MLVBench are presented in Table~\ref{dataset_stats}.

\begin{table}
  \caption{Dataset Statistics}
\centering
\label{dataset_stats}
\resizebox{0.8\linewidth}{!}{
\begin{tblr}{
  cells = {c},
  hlines = {dashed},
  hline{1-2,7} = {-}{solid},
}
                                            & {PyP4LLMSec \\Python} & {VJBench \\ Java} & {CodeQL\\  C++} & {CodeQL \\ Ruby} \\
{Selected\\ Vulnerability}                  & 7          & 12      & 7               & 7                \\
{Original\\ Vulnerability}                  & 7          & 12      & 7               & 7                \\
{Renamed\\ Vulnerability}                   & 7          & 12      & 7               & 7                \\
{Restructured\\ Vulnerability}              & 5          & 12      & 2               & 1                \\
{Rename and\\ Restructured\\ Vulnerability} & 5          & 12      & 2               & 1                
\end{tblr}
}
\end{table}

\begin{figure}
    \centering
    \includegraphics[width=\linewidth]{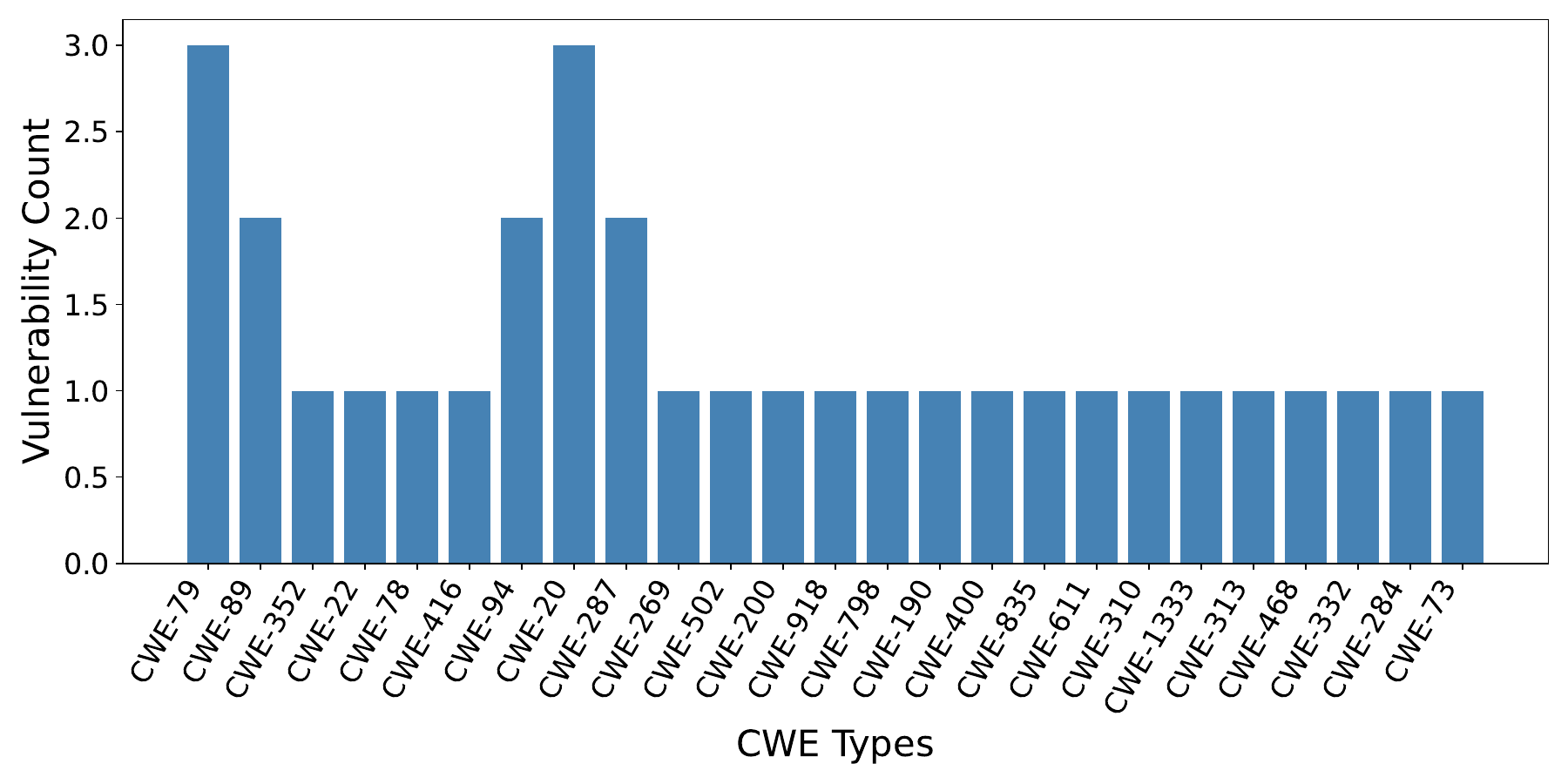}
    \caption{Distribution of Vulnerabilities by CWE Types}
    \label{fig:CWE_types}
\end{figure}

To better align the dataset with real-world security evaluation scenarios, the selection of vulnerabilities from the three source benchmarks, PyP4LLMSec, VJBench, and CodeQL security examples, is guided by the 2024 CWE Top 25 Most Dangerous Software Weaknesses list\footnote{\url{https://cwe.mitre.org/top25/archive/2024/2024_cwe_top25.html}}. However, not all of the Top 25 CWE types are covered in the original datasets. To ensure complete coverage, we supplement the dataset with additional vulnerabilities corresponding to other commonly occurring CWE types, resulting in 33 vulnerabilities spanning 25 distinct CWE categories. The distribution of the selected vulnerabilities and their associated CWE types is presented in Figure~\ref{fig:CWE_types}.

With the selected 33 vulnerabilities, we apply the transformation methodology introduced in VJBench\cite{Effective2023CodeTransform} to generate structurally diverse yet semantically equivalent variants for evaluation. This methodology defines a set of transformation strategies applied to vulnerable functions, including: identifier renaming, if-condition flipping, loop transformation, conditional-statement rewriting, function chaining, function-argument modification, and code-order reordering.
These transformations enhance structural diversity while preserving functional correctness, thereby enabling more robust and cross-language evaluation. Illustrative examples of the applied code transformations are provided in the Appendix, which is hosted on our GitHub repository\footnote{\url{https://github.com/Hahappyppy2024/CFCEval/blob/main/README.md}}.

Following the transformation process, our dataset includes 33 original function-level vulnerabilities along with 106 transformed variants, spanning four programming languages, Python, Java, Ruby, and C/C++, and providing a structurally diverse benchmark for evaluation.

\subsection{Evaluation Dimensions}
\label{dimensions}

Before detailing the specific evaluation dimensions, we first define the standardized format of each evaluation instance used throughout CFCEval.

\subsubsection{\textbf{Evaluation Instance Format}}
\label{setting-def}
\begin{figure}[t]
    \centering
\includegraphics[width=1\linewidth]{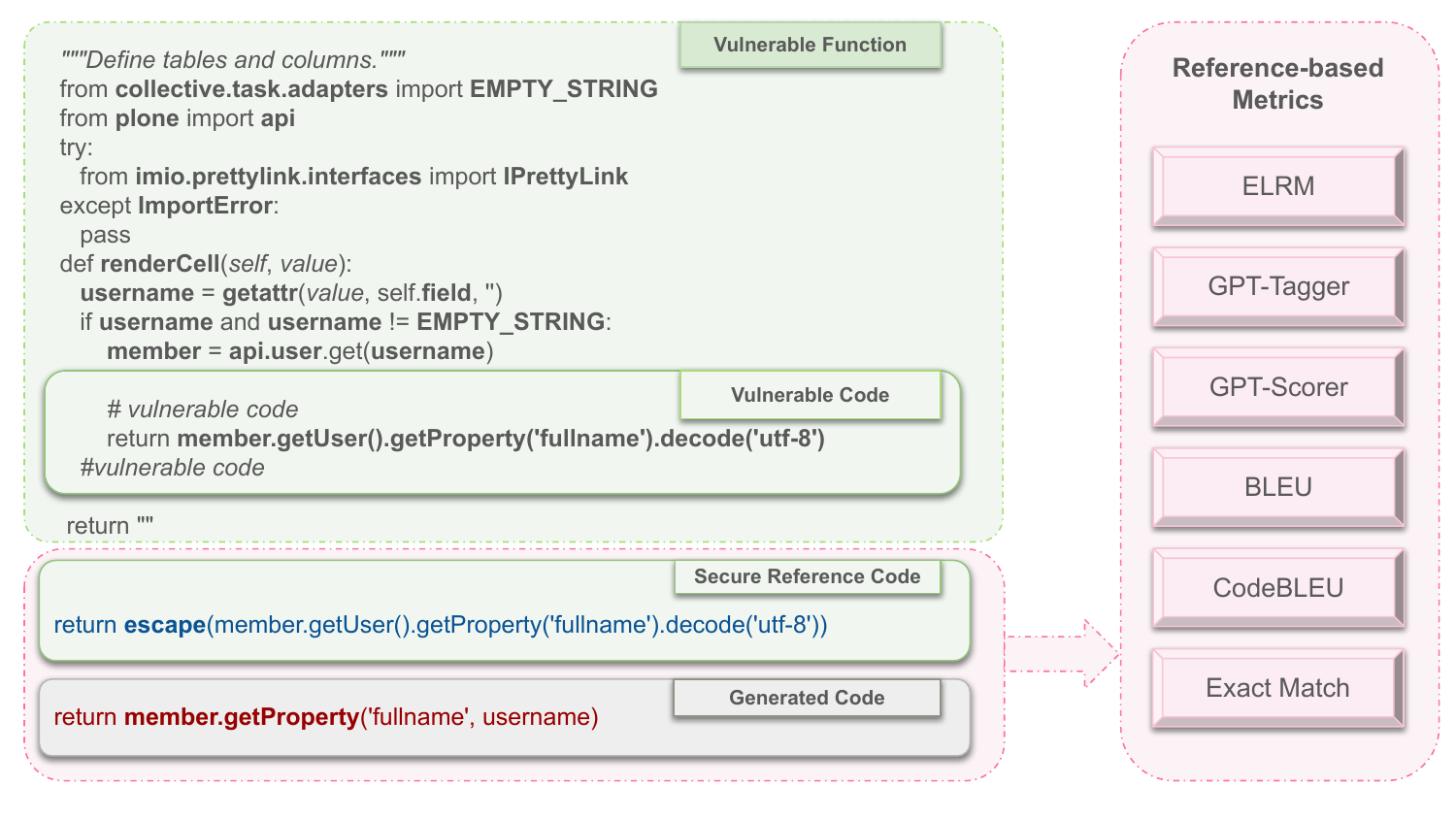}
\caption{The application of the CFCEval framework in our experiments, including vulnerable functions, vulnerable code, secure reference code, and generated code, illustrating the evaluation process.}
\label{entire_procces}
\end{figure}

Each instance for evaluating Code LLMs consists of four key components: the vulnerable function, the vulnerable code (referred to as the vulnerability) within that function, the generated code intended to fix the vulnerability, and the reference code that successfully addresses the vulnerability. These components are represented as a tuple $(F,C_v,C_g,C_r)$ and are illustrated in Figure \ref{entire_procces}, where:

$F$ denotes the \textbf{vulnerable function} that contains the vulnerable code, sourced from a selected security benchmark. 

$C_v$ denotes the \textbf{vulnerable code} located within $F$, identified by a commit message.

$C_g$ denotes the \textbf{generated code} produced by the evaluated Code LLMs based on the prompt $F$.

$C_r$ denotes the secure \textbf{reference code} provided in the fix commit, which successfully repairs $C_v$ in $F$.

Figure~\ref{entire_procces} illustrates the overall structure of the evaluation instance.

\subsubsection{ \textbf{Overview of Evaluation Dimensions}}

CFCEval evaluates each generated output \({C}_g\) across three complementary dimensions, focusing on both code quality and security relevance. An output is considered vulnerability-free only if it satisfies all four dimensions. Representative examples for each category in each dimension are provided in the Appendix of the GitHub repository\footnote{\url{https://github.com/Hahappyppy2024/CFCEval}}.

\textbf{Programming Language Quality (PLanQul.)  }
This dimension identifies and eliminates poorly generated code, specifically \({C}_g\) with syntax errors such as unbalanced brackets or quotes, incorrect starting and ending characters, or invalid terminal characters. Each \({C}_g\) is evaluated and classified as poor or good quality. Instances in which no code is generated are automatically marked as failing this dimension. This dimension can be measured by both reference-free metrics, such as GPT-Tagger and GPT-Scorer, and reference-based metrics, including Exact Match, GPT-Tagger and GPT-Scorer.

\textbf{Fixing Capability (FixCap.) }  
This dimension evaluates the ability of \({C}_g\) to repair the  identified \({C}_v\) that appeared in $F$. It is a critical aspect of program repair in software engineering, commonly used to assess the effectiveness of automatic program repair (APR) tools \cite{lin2007AutoPag, drain2022deepdebug} and Code LLMs \cite{fu2024security, Li2024TestTools}.  We further classify the \({C}_g\) into two distinct categories:

\begin{enumerate}
\item \textit{Fixed: }The \({C}_g\) produced by an individual Code LLM for prompt $F$ successfully fix $C_v$. 
\item \textit{Not Fixed: }The \({C}_g\) produced by an individual Code LLM for prompt $F$ cannot fix  $C_v$ in $F$.
\end{enumerate}
This dimension can also be measured by both reference-free metrics, such as GPT-Tagger and GPT-Scorer, and reference-based metrics, including Exact Match, GPT-Tagger, and GPT-Scorer.

% \vspace{-1.5em}

 \textbf{Post-Transformation Fixing Capability (PTFixCap.) } 
In evaluating \({C}_g\) produced from the transformed function prompt, the original tuple  $(F,C_v,C_g,C_r)$ is updated to $(F_t,C_{vt},C_g,C_{rt})$, where the subscript $t$ denotes the transformed status. This dimension evaluates the ability of \({C}_g\) to repair $C_{vt}$ after applying code transformation rules to $F$. Code transformation is a strategy designed to reduce overlap between Code LLM training datasets and the evaluation dataset, as discussed in recent studies \cite{Thong2022AutoTransform, Effective2023CodeTransform}. The transformation rules, such as identifier renaming and structure modification, are detailed with illustrative examples in Appendix on GitHub. Similarly, \({C}_g\) is classified into two distinct categories: 

\begin{enumerate}
\item \textit{Resolved: }The ${C_g}$ produced by an individual Code LLM for prompt $F_t$ successfully resolves $C_{vt}$.
\item \textit{Unresolved: }The ${C_g}$ produced by an individual Code LLM for prompt $F_t$ fails to resolve $C_{vt}$. 
\end{enumerate}

This dimension can also be evaluated using both reference-free metrics, such as GPT-Tagger and GPT-Scorer, as well as reference-based metrics, including Exact Match, GPT-Tagger, and GPT-Scorer.

\textbf{Element-Level Relevance (ELeRelv.)  } 
Element-Level Relevance is defined as the relevance between \({C}_g\) and \(C_r\) or \(C_{rt}\). This dimension measures these similarities and provides insight into the evolutionary process through which a standard Code LLM transitions into a secure Code LLM. 
Note that this dimension focuses on \({C}_g\), which fails to fix vulnerabilities in $F$ or resolve the transformed vulnerabilities in $F_t$. We classify such \({C}_g\)into the following two categories:

\begin{enumerate}
\item \textit{Relevant: }The \({C}_g\) produced by an individual Code LLM for prompt $F$ or \(F_{v}\), is relevant to \(C_r\) or \(C_{rt}\). 
\item \textit{Irrelevant: }The \({C}_g\) produced by an individual Code LLM for prompt $F$ or \(F_{v}\) is not relevant to \(C_r\) or \(C_{rt}\). 
\end{enumerate}

This dimension can be measured using both reference-free metrics, such as GPT-Tagger and GPT-Scorer, and reference-based metrics, including BLEU, CodeBLEU, GPT-Tagger, and GPT-Scorer. Additionally, this dimension can be evaluated using a novel reference-based metric, termed the Element-Level Relevance Metric (ELRM), which is introduced in the following section.

\subsection{ELRM: An Element-Level Relevance Metric}
During our experiments, we observe that only a small portion of both the original and transformed vulnerabilities could be accurately repaired by existing code LLMs. This raises a fundamental question: How relevant is the generated code to the actual repair patches?
However, current evaluation metrics suffer from significant limitations, failing to capture essential dimensions of code quality and robustness. Widely used metrics such as BLEU and CodeBLEU primarily measure surface-level similarity based on n-gram or lexical overlap, offering limited insight into the structural and semantic correctness of the generated outputs. Specifically, these metrics fall short in the following areas:
(1) Imprecise tokenization: Code lacking whitespace (e.g., x==1) is often treated as a single token, instead of being parsed into meaningful units such as identifiers and operators.
(2) Structural limitations: Short code fragments often cannot be parsed into valid Abstract Syntax Trees (ASTs) or data-flow graphs, limiting the applicability of structure-aware metrics.
(3) Low reference diversity: Most metrics rely on a single ground-truth reference, despite the fact that many repair tasks admit multiple semantically correct predictions. For example, both validators.length(max=256, message=\_('Username too long.')) and validators.length(max=256, message=\_('Username is too long.')) are semantically valid repairs, but existing metrics often fail to recognize this diversity.

These limitations substantially undermine the discriminative power of current metrics in real-world repair evaluation. 
We therefore introduce the Element-Level Relevance Metric (ELRM), which scores fine-grained alignment between a candidate patch and references across identifiers, operators, literals, API invocations, control keywords, and security cues, while allowing multiple valid fixes. 
We next formalize the metric.

\subsubsection{Metric Formulation}
The Element-Level Relevance Metric (ELRM) quantifies the lexical and semantic alignment between generated code and reference implementations through a weighted combination of complementary sub-metrics:

\begin{multline}
\Large
\text{ELRM} = \alpha \cdot \text{BLEU} + \beta \cdot \text{BLEU}_{\text{weight}} \\
\Large
+ \lambda \cdot \text{BLEU}_{ \text{keywords\_ops}} + \mu \cdot \text{Similarity}_{ \text{string\_literal}}
\end{multline}

where BLEU is a standard metric for evaluating the quality of machine-generated code by comparing n-gram overlaps with reference code~\cite{bleu2022Papi}.

$\text{BLEU}_{\text{weight}}$ is a weighted variant of BLEU that assigns different importance to keywords and non-keywords in both the reference and the generated output.

$\text{BLEU}_{\text{keywords\_ops}}$ computes the BLEU score specifically over sequences of programming language keywords and operators.

$\text{Similarity}_{\text{string\_literal}}$ measures the lexical similarity of string literals between the reference and the generated code.

This formulation extends CodeBLEU by replacing its AST- and data-flow-based components 
with more lightweight yet effective lexical-level signals: 
$\text{BLEU}_{\text{keywords\_ops}}$ and $\text{Similarity}_{\text{String\_literal}}$, making it more suitable for evaluating short, structure-sparse code generations, where AST and data-flow information may be sparse or unreliable. The following items provide detailed descriptions of each component of the formulation.

\begin{enumerate}
    \item \textbf{N-Gram Match and Weighted N-Gram Match} \\
The standard BLEU metric measures n-gram overlap between generated and reference code. However, its effectiveness in natural language tasks does not transfer well to programming languages, where strict syntax and limited keywords (e.g., \textit{int}, \textit{public}, \textit{=}, \textit{+}) make vanilla BLEU insensitive to the functional importance of language-specific tokens.
To account for code-specific structure, \textbf{CodeBLEU}~\cite{ren2020codebleu} employs a weighted n-gram scheme that emphasizes programming keywords by assigning them higher weights (e.g., 1). The weighted precision is computed as:
\begin{equation}
p_n = \frac{ \sum\limits_{C \in \text{Candidates}} \sum\limits_{i=1}^{l} \mu_n^i \cdot \text{Count}_{\text{clip}}(C(i, i+n)) }{ \sum\limits_{C' \in \text{Candidates}} \sum\limits_{i=1}^{l} \mu_n^i \cdot \text{Count}(C'(i, i+n)) }
\end{equation}

where $C(i, i+n)$ denotes the n-gram starting at position $i$ with length $n$, and $\mu_n^i$ is the assigned weight of the $i$-th n-gram as CodeBLEU~\cite{ren2020codebleu}  presents.

To penalize overly short candidates, CodeBLEU incorporates the standard BLEU brevity penalty \cite{bleu2022Papi}:
\begin{equation}
\text{BP} =
\begin{cases}
1 & \text{if } c > r \\
e^{1 - r/c} & \text{if } c \leq r
\end{cases}
\end{equation}

where $c$ stands for the length of the candidate and $r$ is the effective reference length as BLEU~\cite{bleu2022Papi} presents.

By incorporating both weighted precision and the brevity penalty, the final weighted BLEU score is then calculated as:
\begin{equation}
\text{BLEU}_{\text{weight}} = \text{BP} \cdot \exp\left( \sum_{n=1}^{N} w_n \log p_n \right)
\end{equation}

We briefly restate the core formulations from CodeBLEU~\cite{ren2020codebleu} and BLEU~\cite{bleu2022Papi} to clarify the underlying design of its weight-ed n-gram precision and brevity penalty.
In our implementation, we further refine the tokenization procedure to address scenarios where the absence of whitespace causes entire lines of code to be interpreted as single tokens. This refinement ensures the correct separation and identification of key syntactic elements such as identifiers, keywords, and operators, thereby improving the fidelity of n-gram matching in code evaluation.

\item \textbf{Fine-Grained Tokenization}\\
To address the coarse-grained tokenization limitations in CodeBLEU, we introduce a refined tokenization strategy that more accurately reflects the syntactic granularity of source code. For example, CodeBLEU treats the expression \textit{permission\_classes=[permissions.IsAuthenticated, IsSuperUser]} as a single token, thereby failing to distinguish critical lexical components such as identifiers, keywords, operators, and delimiters. In contrast, our approach tokenizes the same expression into a list of fine-grained tokens:
\textit{["permission\_classes", "=", "[", "permissions", ".", "IsAuthenticated", ",", "IsSuperUser", "]"]}
This fine-grained tokenization facilitates more accurate structural alignment between the generated and reference code, which is essential for evaluating semantic and syntactic relevance in program synthesis.

\item \textbf{Language Keywords and Operators Match}\\
Unlike CodeBLEU, which incorporates AST and data-flow comparisons, our method focuses exclusively on the sequences of programming language keywords and operators. This design choice stems from the observation that AST- and data-flow-based comparisons are often unreliable or infeasible when evaluating short code snippets, where structural information is limited or absent.
Concretely, we extract the keyword and operator sequences (e.g., for, if, return, =, +) from both the candidate and reference implementations, and compute their n-gram overlap using the BLEU metric to assess lexical alignment.

\item \textbf{String Literal Similarity}\\
In many cases, two code snippets may be semantically and syntactically equivalent while differing only in their string literals, which refer to constant string values explicitly defined within the source code. Such differences, although functionally irrelevant, can significantly affect string-based similarity metrics, especially in short code snippets. For example, the statement \textit{setMessage("Invalid input format")} contains the string literal \textit{"Invalid input format"}, which could be rewritten as \textit{setMessage("Input format is invalid")}. Although the surrounding logic remains unchanged, such differences can affect the evaluations results, as illustrated in the example shown in Figure~\ref{fig:ELRM_eg2}. To capture these subtle lexical variations, we compute the similarity between string literals using three complementary syntactic metrics :
(1)~\textbf{Levenshtein distance}~\cite{levenshtein1966binary}, which quantifies character-level edit operations;
(2)~\textbf{SequenceMatcher}, which identifies the longest common subsequences; and 
(3)~\textbf{Jaccard similarity}~\cite{jaccard1912distribution}, which captures token overlap from a set-theoretic perspective. 
The final score is calculated as their average, leveraging the 
complementary strengths of all three for robust string-level alignment.

\end{enumerate}

\subsubsection{Two Examples and Analysis}
In this section, we present two simple examples to demonstrate the calculation of ELRM. Additionally, we highlight the qualitative benefits of ELRM in comparison to the CodeBLEU metric.
\paragraph{ \textbf{Example 1} }
\begin{figure}[H]
    \centering
    \includegraphics[width=1\linewidth]{0images/example1.pdf}
    \caption{Example 1. CodeBLEU: 22.03 , ELRM: 13.71.}
    \label{fig:ELRM_eg1}
\end{figure} 
Figure~\ref{fig:ELRM_eg1} illustrates a generated code snippet and its corresponding reference code. In this example, the key difference lies in the presence of an unnecessary segment,\textit{get\_ipython\_package\_dir()}, in the generated output (i.e., the candidate in BLEU evaluation), which is absent from the reference code. This deviation is expected to be effectively penalized by a robust evaluation metric.

To compute the ELRM score (normalized to a 0-100 scale), we follow four sequential steps:
(1)~\textbf{N-Gram Match:} The standard BLEU score is computed to quantify the n-gram overlap between the candidate and the reference, yielding a score of 25.27.
(2)~\textbf{Weighted N-Gram Match:} Since the code line contains no programming language keywords, each token is assigned an equal weight of 0.2. The resulting weighted BLEU score is 70.7.
(3)~\textbf{Keyword and Operator Match:} The keywords in the reference code are \textit{["=", "[", "]"]}, while those in the generated code are \textit{["=", "[", ",", "(", ")", "]"]}. Programming keywords and operators are extracted as ordered sequences and evaluated using the BLEU metric, resulting in a score of 9.55. 
(4)~\textbf{String Literal Similarity:} As neither the reference nor the candidate contains string literals, the similarity score for this component is 0.
By applying the combination weights $\alpha = 0.10$, $\beta = 0.05$, $\lambda = 0.80$, and $\mu = 0.05$, the final ELRM score is computed to be 13.71, which is substantially lower than the corresponding CodeBLEU score in this case.

\paragraph{\textbf{Example 2}}
\begin{figure}[H]
\centering
\includegraphics[width=1\linewidth]{0images/example2.pdf}
\caption{Example 2. CodeBLEU: 62.87, ELRM: 94.95.}
\label{fig:ELRM_eg2}
\end{figure}
To facilitate a comparative analysis between ELRM and CodeBLEU,
Figure~\ref{fig:ELRM_eg2} presents a second example comprising a generated code snippet and its corresponding reference implementation.
In this example, the only difference lies in the use of double quotes (i.e., \textit{"}) versus single quotes (i.e., \textit{'}).
Despite this minor variation, the candidate and the reference are semantically and functionally equivalent.
Consequently, the metric is expected to yield a high score in this case, reflecting the negligible discrepancy.
To evaluate the ELRM score (on a normalized 0-100 scale), we execute the following four-step procedures:
To evaluate the ELRM score (normalized to a 0-100 scale), we follow a four-step procedure:
(1)~\textbf{N-Gram Match:} We begin by computing the standard BLEU score to assess n-gram overlap between the generated output and its reference. In this example, the BLEU score is 64.07.
(2)~\textbf{Weighted N-Gram Match:} For weighted evaluation, programming keywords are assigned a weight of 1.0, while all other tokens receive a uniform weight of 0.2. This adjustment yields a weighted BLEU score of 70.89.
(3)~\textbf{Keyword and Operator Match:} We extract the following ordered list of programming-specific keywords and operators from both the candidate and the reference: ["if", "==", "and", ".", "(", ")", ":"]. The BLEU score computed over this sequence is 100.0.
(4)~\textbf{String Literal Similarity:} As both the candidate and reference contain the identical string literal \textit{"add"}, this component contributes a similarity score of 100.0.
Using the weighting configuration $\alpha = 0.10$, $\beta = 0.05$, $\lambda = 0.80$, and $\mu = 0.05$, the final ELRM score is computed to be 94.95, which is higher than the corresponding CodeBLEU score for this example.

\paragraph{\textbf{Summary}}
The two examples demonstrate that ELRM enables finer-grained and more faithful evaluation than CodeBLEU. 
It effectively addresses the three core deficiencies of existing metrics: (1) by applying element-aware tokenization, it resolves the imprecise handling of compact code; (2) by operating on typed lexical units rather than full ASTs, it remains robust on short code fragments that cannot be structurally parsed; and (3) by supporting multiple reference variants, it mitigates low reference diversity and recognizes semantically equivalent fixes. 
Consequently, ELRM penalizes semantically redundant outputs (Example1) and rewards functionally equivalent variants with minor syntactic differences (Example2), capturing both syntactic structure and functional meaning. 
The next section presents empirical results validating its effectiveness and robustness across benchmarks.

\section{Experimental Setup and Results}
\label{experiment}
\subsection{Experimental Setup}
We evaluate four representative code LLMs, consistent with those used in the original CodeBLEU evaluation: \textbf{GitHub Copilot}\footnote{\url{https://code.visualstudio.com/docs/copilot/overview}}, \textbf{Cursor}\footnote{\url{https://cursor.com/cn/agents}} (Gemini-2.5-flash), \textbf{DeepCoder}\footnote{\url{https://ollama.com/library/deepcoder:14b}}, and \textbf{CodeGeeX4}\footnote{\url{https://github.com/THUDM/CodeGeeX}}. All models are assessed using their latest available versions. Evaluations are conducted on \textbf{MLVulBench}, a multi-language vulnerability benchmark within CFCEval, featuring prompts in Python, Java, C++, and Ruby.

\subsection{Results}

Table~\ref{tab:baselines} reports the ELRM, BLEU, and CodeBLEU scores, along with LLM-based evaluations, on 20 randomly sampled benchmark cases spanning all transformation types and four programming languages. Each input is paired with four generated outputs, resulting in 80 input-output prompt pairs for LLM-based scoring. The first three metrics are normalized to a 0-100 range, while LLM-based scores use a 5-point Likert scale (1=very poor, 5=excellent). Example prompts are provided in the Appendix of our GitHub repository \footnote{\url{https://github.com/Hahappyppy2024/CFCEval/blob/main/README.md}}. Based on these results, we investigate the following four research questions.

\begin{table}[h]
\caption{
The results of all baselines on the given vulnerabilities, evaluated by ELRM, BLEU, CodeBLEU, and LLM-based evaluation scores.
}
\captionsetup{skip=2pt}
\centering
\label{tab:baselines}
\begin{tblr}{
  column{even} = {c},
  column{1} = {c},
  column{3} = {c},
  hlines = {dashed},
  hline{1-2,6} = {-}{solid,0.08em},
  rowsep=1pt,
}
                   & ELRM  & BLEU  & CodeBLEU & LLMs \\
Cursor             & 24.72 & 30.33 & 29.98    & 2.38 \\
GitHub Copilot & 22.93 & 29.43 & 29.43    & 2.9  \\
CodeGeeX4          & 29.19 & 36.22 & 30.75    & 2.4  \\
DeepSeekCoder          & 18.65 & 21.19 & 24.16    & 1.92 
\end{tblr}
\end{table}

\subsubsection{\textbf{RQ1}} To what extent is ELRM a reliable and stable metric for differentiating code generation models, in terms of score significance and variance?\textit{\\}

We compute the ELRM scores of the generated code from all selected Code LLMs, and calculate their means, variances, and paired \textit{t}-statistics, which are reported in Table~\ref{tab:baselines} and Table~\ref{tab:mean_std_t}. From Table~\ref{tab:baselines}, we observe that Cursor, Copilot, and CodeGeeX exhibit relatively close ELRM scores (means ranging from 0.229 to 0.292), while DeepSeekCoder obtains a substantially lower mean of 0.186. This is consistent with our qualitative observation that DeepSeekCoder, as a locally deployed model, frequently produces outputs with syntactic and readability issues, whereas the other systems generate more consistent and well-formed code.

\begin{table}[h]
\caption{
Mean, standard deviation, and paired \textit{t}-statistics of ELRM across Code LLMs. Each \textit{t}-statistic compares a system with the one above; the first compares Cursor with DeepSeek-Coder.
}
\centering
\label{tab:mean_std_t}
\begin{tblr}{
  cells = {c},
  hlines = {dashed},
  hline{1-2,6} = {-}{solid,0.08em},
}
                   & Mean  & StdDev & t    \\
Cursor             & 24.72 & 23.75  & 1.86 \\
GitHub Copilot & 22.93 & 21.95  & 0.54 \\
CodeGeeX4          & 29.20 & 25.87  & 1.66 \\
DeepSeek-Coder          & 18.65 & 20.36  & 2.81 
\end{tblr}
\end{table}
In Table~\ref{tab:mean_std_t}, each t-statistic compares a system with the one above it. For instance, the first t-statistic compares Cursor with DeepSeek-Coder. The paired t-statistic between CodeGeeX and DeepSeek-Coder reaches 2.81 ($p = 0.0056$), indicating a statistically significant difference. While other pairwise comparisons show less statistical significance (e.g., $p = 0.10$ and $p = 0.0651$), they are directionally consistent with the overall system quality and fall within the sensitivity threshold observed in LLM-based annotations. These results suggest that while the differences are not always statistically significant, they align with the general trends in model performance.
The standard deviations (20.36-25.87) reflect expected variability across diverse prompts and model behaviors. Despite this variability, ELRM maintains consistent scoring trends, demonstrating its robustness to real-world input diversity. Notably, the pairwise t-test between CodeGeeX4 and DeepSeek-Coder yields a high statistic (t = 2.81), indicating a statistically meaningful performance gap that ELRM is able to capture. Even in cases with marginal significance (e.g., Cursor vs. Copilot, t = 0.54), the score differences align directionally with overall system quality. These results suggest that ELRM can effectively discriminate between higher- and lower-quality outputs. We therefore conclude that ELRM provides a reliable and valid basis for evaluating code generation performance. Additional ablation studies are available in the Appendix and on our GitHub repository\footnote{\url{https://github.com/Hahappyppy2024/CFCEval/blob/main/README.md}}.

\subsubsection{\textbf{RQ2}} Does ELRM achieve higher correlation with LLMs judgments than CodeBLEU and BLEU in evaluating generated code?\textit{\\}

\begin{table}[h]
\centering
\caption{
Comparison of the Pearson correlation coefficients between LLM-based evaluation scores and different metrics.
}
\label{tab:pearson}
\begin{tblr}{
  cells = {c},
  hlines = {dashed},
  hline{1,2,6} = {-}{solid,0.08em},
  rowsep=1pt, % 行间距变小
  colsep=2pt, % 列间距紧凑
}
Metrics                          & {Pearson's Correlation} \\
BLEU (Orignal) \textsuperscript{a}      & -0.173                        \\
CodeBLEU                  & 0.461                         \\
BLEU (ELRM) \textsuperscript{b}     & 0.804                         \\
ELRM                        & 0.816                         
\end{tblr}
\parbox{0.95\linewidth}{\footnotesize
BLEU\textsuperscript{a} uses the CodeBLEU tokenizer; 
BLEU\textsuperscript{b} uses the ELRM tokenizer.
}
\end{table}

Owing to their extensive pretraining on large-scale code corpora, large language models (LLMs) have demonstrated exceptional performance on function-level tasks, often outperforming human coders in terms of accuracy and efficiency~\cite{chen2021evaluating, khan2023assessingpromisepitfallschatgpt}. LLMs have shown to be highly proficient in generating function-level code with an understanding of both syntax and semantics, which enables them to produce reliable outputs for complex tasks. Furthermore, the labels generated by LLMs tend to align more closely with gold-standard annotations compared to those produced by human annotators, showcasing their ability to consistently reproduce high-quality outputs~\cite{parfenova-etal-2025-text}. Consequently, we adopt labels generated by three distinct LLM versions—ChatGPT-4o, ChatGPT-o3, and ChatGPT-4.1—as the reference judgments for evaluating 20 selected vulnerabilities from the MLVulBench benchmark.

As demonstrated in Table~\ref{tab:pearson}, ELRM achieves the highest Pearson correlation with the LLM-generated annotations, indicating that it most effectively aligns with the semantic and functional evaluations provided by the LLMs. This suggests that ELRM is particularly well-suited for capturing the nuanced judgments made by these models. In contrast, CodeBLEU frequently produces zero scores, which limits its ability to accurately capture semantic relevance and suppresses its correlation with LLM-generated ratings. This lack of sensitivity highlights the limitations of CodeBLEU, especially in tasks related to vulnerability detection and repair, where fine-grained semantic matching is crucial. These findings underscore ELRM’s superiority as an evaluation metric, particularly in the context of code generation for vulnerability-related tasks, where the ability to capture both syntax and semantics is paramount.

\subsubsection{\textbf{RQ3}} Does ELRM achieve higher correlation with human judgments than CodeBLEU and BLEU in evaluating generated code? \textit{\\}

To evaluate the effectiveness of the Element-Level Relevance Metric (ELRM), we conduct a comprehensive assessment on the full set of function-level vulnerabilities within the MLVBench benchmark, utilizing four prominent code generation models: Cursor, Copilot, CodeGeeX, and DeepSeek-Coder. Each model is evaluated based on its ability to generate code patches for these vulnerabilities. To ensure consistency and reliability, two independent annotators score the generated patches on a 1-5 scale. Due to the varying performance of these models, sample sizes across models differ: Cursor and Copilot each produce results for 97 functions, CodeGeeX for 77 functions, and DeepSeek-Coder for 91 functions.

\begin{table}[h]
\centering
\caption{Correlations with human judgments on MLVBench (function level). Pearson's $\gamma$ between \textit{HumanAvg} (two 1–5 ratings; Cohen’s $\kappa=0.7663$) and BLEU/CodeBLEU/ELRM per model ($n$ shown). Bold = strongest per model.}
\label{tab:human_judge}
\begin{tblr}{
  cell{1}{1} = {c},
  hlines = {dashed},
  hline{1-2,6} = {-}{solid,0.08em},
}
Model &  n & BLEU & CodeBLEU & ELRM   \\
Cursor         & 97         & 0.3548        & 0.2253            & \textbf{0.8281} \\
Copilot        & 97         & 0.3975        & 0.3096            & \textbf{0.6681} \\
CodeGeeX       & 77         & 0.2931        & 0.2174            & \textbf{0.8601} \\
DeepSeekCoder  & 91         & 0.0787        & 0.0772            & \textbf{0.7991}
\end{tblr}
\end{table}

The inter-annotator agreement is robust, with a Cohen’s kappa value of 0.77, indicating substantial consistency between the raters. For each model, the average score, referred to as HumanAvg, is computed by averaging the two independent ratings. Subsequently, Pearson’s correlation coefficient is calculated between HumanAvg and the evaluation scores derived from three different metrics: BLEU, CodeBLEU, and ELRM. This allows for a comparative analysis of the performance of each metric in capturing human judgment.

As shown in Table~\ref{tab:human_judge}, ELRM consistently demonstrates the strongest correlation with human evaluations, outperforming both BLEU and CodeBLEU. These results validate ELRM's ability to provide a more accurate and nuanced measure of code quality, especially in the context of function-level vulnerability repair.

\subsubsection{\textbf{RQ4}} Can CFCEval serve as the first comprehensive benchmark for automatic evaluation metrics that assess code LLMs in terms of both quality and security? \textit{\\}

To use CFCEval as an automatic benchmark, we construct reference-based prompts following the methodology of QUDeval~\cite{wu-etal-2023-qudeval}. We introduce two metrics, both based on ChatGPT-4.1mini: \textbf{GPT-Tagger}, which classifies the generated code according to CFCEval's dimensions, and \textbf{GPT-Scorer}, which rates code quality on a 5-point scale. We evaluate the code generated by GitHub Copilot and CodeGeeX for 20 MLVBench vulnerabilities. Example prompts are provided in the Appendix on GitHub repository \footnote{\url{https://github.com/Hahappyppy2024/CFCEval/blob/main/README.md\#gpt-based-metrics-reference-based-prompts}}.

\begin{table}[ht]
\centering
\caption{Reference-based assessment of code generated by GitHub Copilot and CodeGeeX using GPT-Tagger and GPT-Score across PLanQu., FixCap., PTFixCap., and EleReLv metrics.}
\label{tab:GPT-ts}
\begin{tblr}{
  colsep = 4pt,  % Reduce column separation
  row{1} = {c},
  row{2} = {c},
  row{3} = {c},
  row{4} = {c},
  row{5} = {c},
  row{6} = {c},
  row{9} = {c},
  row{10} = {c},
  row{11} = {c},
  row{12} = {c},
  cell{1}{3} = {c=2}{},
  cell{1}{5} = {c=2}{},
  cell{3}{1} = {r=2}{},
  cell{5}{1} = {r=2}{},
  cell{7}{3} = {c=2}{c},
  cell{7}{5} = {c=2}{c},
  cell{8}{3} = {c},
  cell{8}{4} = {c},
  cell{8}{5} = {c},
  cell{8}{6} = {c},
  cell{9}{1} = {r=2}{},
  cell{11}{1} = {r=2}{},
  hline{1,3,5,7,9,11,13} = {-}{},
  hline{2,8} = {3-6}{},
  hline{4,6,10,12} = {2-6}{dashed},
}
         &            & PLanQul. (\%)  &        & FixCap. (\%)  &       \\
         &            & Poor           & Good   & Not Fixed     & Fixed \\
Copilot  & GPT-Tagger & 60.0           & 40.0   & 75.0          & 25.0  \\
         & GPT-Score  & 60.0           & 40.0   & 70.0          & 30.0  \\
CodeGeeX & GPT-Tagger & 55.0           & 45.0   & 85.0          & 15.0  \\
         & GPT-Score  & 80.0           & 20.0   & 90.0          & 10.0  \\
         &            & PTFixCap. (\%) &        & ELeReLv. (\%) &       \\
         &            & UnRes.         & Res.   & Irre.         & Rel.  \\
Copilot  & GPT-Tagger & 75.0           & 25.0   & 70.0          & 30.0  \\
         & GPT-Score  & 70.0           & 30.0   & 55.0          & 45.0  \\
CodeGeeX & GPT-Tagger & 80.0           & 20.0   & 80.0          & 20.0  \\
         & GPT-Score  & 90.0           & 10.0   & 90.0          & 10.0  
\end{tblr}
\end{table}

% All dimensions in CFCEval can be evaluated with Exact Match~\cite{ding2023crosscodeeval} and ELRM against the reference code. 
As shown in Table~\ref{tab:GPT-ts}, CFCEval evaluates four dimensions spanning code quality and security, using both quantitative metrics and LLM-based scoring prompts. The table reports LLM evaluation results based on reference-based prompts designed to assess each dimension. For instance, in the PLanQu. dimension, GitHub Copilot's GPT-Tagger scores 60\% poor and 40\% good, while CodeGeeX scores 55\% poor and 45\% good. In FixCap., Copilot's GPT-Score evaluates 75\% as not fixed and 25\% as fixed (out of 20 vulnerabilities), while CodeGeeX scores 85\% and 15\%, respectively. These examples illustrate CFCEval's ability to assess the code quality and security using reference-based prompts. While our current evaluation focuses on the reference-based setting, CFCEval also supports reference-free metrics, with corresponding prompts provided in the Appendix on GitHub repository.

\section{Pilot Study: Evaluation of Code LLMs for Vulnerability Repair}
\label{study}

We conduct a pilot study comparing two prominent code generation LLMs—Cursor and Copilot—on real-world benchmark instances, represented as a tuple $(F, C_v, C_g, C_r)$. For each instance, $C_v$ denotes the vulnerable code snippet, $C_r$ is the secure reference patch, and $C_g$ is the model-generated fix. Both models are evaluated under identical prompts and decoding settings, generating candidate patches for the given vulnerabilities. To assess robustness, we apply a single program transformation, identifier renaming, to obtain the transformed instance $F_t$, and re-evaluate the model-generated patch $C_{g}$ and the reference patch $C_{rt}$.

In this pilot study, we use several metrics from the CFCEval framework to assess the quality of the generated fixes. FixCap and PTFixCap apply Exact Match with $C_r$ and $C_{rt}$, respectively, to evaluate whether the generated patch successfully eliminates the vulnerability. FixCap checks if $C_g$ exactly matches $C_r$, ensuring the correctness of the fix, while PTFixCap evaluates the robustness of the fix under distribution shift by comparing $C_{g}$ against $C_{rt}$. Element-Level Relevance (ELeRelv) is measured using the Element-Level Relevance Metric (ELRM), which quantifies the fine-grained relevance between non-matching elements, providing a more detailed analysis of the semantic alignment of the generated patch with the reference patch.

This pilot study serves as a preliminary evaluation of the Code LLMs' performance in vulnerability repair tasks. It offers insights into their ability to generate accurate and robust fixes. Further details, including experimental setups and additional results, are available in the supplementary document provided at \href{https://github.com/Hahappyppy2024/CFCEval/blob/main/Pilot%20Case%20Study.pdf}{Supplementary Document}.

\section{Threats to Validity}
\label{limit}

\textit{Two human judges.} Human evaluations are prone to subjectivity, expertise variance, and fatigue. To mitigate these, we standardize the evaluation process using a shared rubric, implement blinding and randomization, and apply double-annotation for each code sample. We measure inter-rater reliability using Cohen's $\kappa$ and resolve disagreements through adjudication. Additionally, spot-checking with tests and static analysis ensures that the evaluations align with expected outcomes and maintain consistency. This approach helps minimize bias and ensures the robustness of human judgments in complex evaluation tasks.

\textit{LLM-as-judge (ChatGPTs).} LLM-based evaluations can be affected by model biases and variations in prompts and versions. To address this, we standardize inputs, randomize A/B testing, and fix prompts to minimize bias. We log model IDs and dates to track changes and ensure transparency in the evaluation process. Multiple responses are aggregated to reduce outlier effects, reflecting the model's judgment rather than inconsistencies in prompt or version. This strategy ensures that results represent the model's capabilities while minimizing errors introduced by prompt-specific biases.

\section{Related Work}
\label{related_work}
Recent research has explored the use of large language models (LLMs) for various code generation tasks, including program repair, debugging, and vulnerability detection. Early efforts in automatic program repair (APR) demonstrated the potential of LLMs to generate code fixes for common errors, but these methods were limited by small datasets and handcrafted rules~\cite{lin2007AutoPag, drain2022deepdebug}. As LLMs have evolved, their application has expanded to security-related tasks, such as vulnerability detection and repair, though challenges remain in adapting general-purpose models to address the subtleties of security-critical tasks~\cite{fu2024security, Li2024TestTools, securityForge, securityEMSE}. These challenges stem from the complexity of security flaws, which often require nuanced understanding and precise fixes beyond simple syntactic corrections.

Common benchmarks like HumanEval~\cite{chen2021evaluating}, ReCode~\cite{wang-etal-2023-recode}, and CoderEval~\cite{Hao2024CoderEval} focus on functional correctness, offering insights into syntactic validity. However, specialized benchmarks such as LLMSecEval~\cite{tony2023llmseceval} and PyP4LLMSec~\cite{cheng2024pyp4llmsec} have emerged to assess LLMs on security-oriented tasks. Despite these efforts, security datasets like DS-1000~\cite{szalontai2024largelanguagemodelscode} still struggle to capture the full complexity of real-world vulnerabilities, particularly those that arise in dynamic software environments. Moreover, existing datasets often fail to account for evolving attack vectors and edge cases that are critical for robust security evaluation.

The evaluation of LLMs has expanded beyond functional correctness to include robustness under perturbations~\cite{robustness}, additional quality metrics~\cite{beyondcorrectness}, and fairness~\cite{fairness}. While metrics like BLEU~\cite{bleu2022Papi} and CodeBLEU~\cite{ren2020codebleu} are widely used to assess syntactic and structural similarities, they fail to capture deeper semantic and security-related aspects critical for vulnerability repair. Metrics like Exact Match (EM)~\cite{ding2023crosscodeeval} focus on syntactic correctness but overlook finer-grained code qualities essential for secure and robust software generation. As a result, there is an increasing demand for new, more comprehensive metrics that address both the functional and security dimensions of code quality.

In summary, while LLMs have made significant strides in security-critical applications, there remains a need for more specialized benchmarks and evaluation methods to address the complex demands of vulnerability detection and repair. Developing semantic, robust, and security-focused metrics will be crucial for enhancing the reliability and security of LLM-generated code in real-world software systems.

\section{Conclusion}
\label{conclusion}

In conclusion, we present the Code Fix Capability Evaluation (CFCEval) framework, which addresses key evaluation limitations by leveraging the MLVBench dataset to mitigate training-testing data overlap bias. CFCEval evaluates code LLMs across four dimensions, including Element-Level Relevance, assessed using the Element-Level Relevance Metric (ELRM). Experimental results show that ELRM better aligns with LLM quality scores compared to existing metrics like CodeBLEU and BLEU. Additionally, GPT-based metrics confirm CFCEval's potential as an automatic evaluation benchmark. The pilot study presented in this paper demonstrates CFCEval's practical application and its ability to evaluate code generation models in real-world vulnerability repair tasks. 
In the future, we plan to incorporate additional vulnerabilities from a wider range of programming languages into the MLVBench dataset and conduct a comprehensive case study to further validate our framework and explore its applicability across diverse software ecosystems.

\bibliographystyle{IEEEtran}
\bibliography{ref}
\end{document}